\documentclass{PoS}
\usepackage{tabularx}
\usepackage{caption}

\newcommand{\beq}{\begin{equation}}
\newcommand{\eeq}{\end{equation}}
\newcommand{\beqa}{\begin{eqnarray}}
\newcommand{\eeqa}{\end{eqnarray}}

\title{Nuclear Chiral EFT in the Precision Era}

\ShortTitle{Nuclear $\chi$EFT in the Precision Era}

\author{\speaker{Evgeny Epelbaum}
\thanks{It is a great pleasure to thank my collaborators Hermann
  Krebs, Ulf-G.~Mei{\ss}ner
and all members of the LENPIC collaboration
for sharing their insights into the discussed topics and the
organizers of CD2015 for making this exciting workshop possible. I am
also grateful to Hermann
  Krebs and Ulf-G.~Mei{\ss}ner for useful suggestions on the
  manuscript. This work was
supported in part by the ERC project 259218 NUCLEAREFT and by DFG
(SFB/TR 16, ''Subnuclear Structure of Matter''). }
\\
        Institut f\"ur Theoretische Physik II, Ruhr-Universit\"at Bochum,
  D-44780 Bochum, Germany\\
        E-mail: \email{evgeny.epelbaum@rub.de}}

\abstract{Chiral effective field theory has established itself as the
  method of choice to study nuclear forces and low-energy nuclear
  dynamics. I review the status and prospects of this approach and discuss
  ongoing efforts to advance the precision frontier for \emph{ab
    initio} description of few-nucleon systems. Special emphasis is
  put on the precise determination of the two-nucleon force at fifth
  order in the chiral expansion, role of the chiral symmetry,
  the convergence pattern of the chiral expansion and
  the quantification of the theoretical uncertainties. The discussed
  topics are essential for ongoing studies towards elucidating the structure
  of the three-nucleon force which will be briefly addressed as well. }

\FullConference{The 8th International Workshop on Chiral Dynamics, CD2015 ***\\
		29 June 2015 - 03 July 2015\\
		Pisa, Italy}

\begin{document}

\section{Introduction}

The classical problem of the nuclear force has experienced a
strong revival since the 1990s when Weinberg has shown that chiral
perturbation theory (ChPT) can, in fact, be fruitfully applied beyond the
Goldstone-boson and single-nucleon sectors
\cite{Weinberg:1990rz,Weinberg:1991um}. 
While a direct
application of ChPT to nucleon-nucleon (NN) scattering amplitude 
makes little sense due to the intrinsically non-perturbative nature of the
problem, Weinberg has realized that all diagrams which violate the chiral
power counting 
emerge from iterations of a Lippmann-Schwinger (LS) type of equation.
The task of calculating the scattering amplitude thus naturally
reduces to the conventional quantum  mechanical $A$-body problem with
nuclear potentials derived in ChPT. The resulting approach is 
commonly referred to as (nuclear) chiral effective field theory (EFT)
and has the appealing features of being model-independent,
systematically improvable and related to QCD via its symmetries, see
Refs.~\cite{Epelbaum:2008ga,Machleidt:2011zz} for recent review articles. 

Today, after a quarter of a century of intense research, nuclear chiral
EFT is entering the precision era and is expected to shed light on the
long-standing unsolved problems such as especially
the three-nucleon force (3NF) problem \cite{KalantarNayestanaki:2011wz,Hammer:2012id}. This is
becoming possible due to advances in the following three
areas: First, the chiral expansion of the nuclear force has been/is
being pushed to a sufficiently high order for making precision
calculations. It is essential that the 
determination of an increasing number of unknown 
low-energy constants (LECs), which often plagues higher-order calculations in
ChPT, is manageable in the nuclear sector thanks to the vast amount of
neutron-proton (np) and proton-proton (pp) scattering data and due to
the fact that most of the unknown LECs contribute to the NN force. Secondly,
chiral EFT provides a natural framework to quantify the theoretical
uncertainties of the calculations, the issue which needs to be
seriously addressed
in order to make a meaningful comparison between theory and
experiment. Finally, recent advances in \emph{ab initio} few- and
many-body methods including the coupled-cluster expansions \cite{Hagen:2012sh},
the no-core shell model \cite{Jurgenson:2013yya}, Green's function Monte Carlo method
\cite{Lovato:2013cua}, the in-medium similarity renormalization group approach \cite{Hergert:2012nb},  
and the self-consistent Green's functions technique \cite{Soma:2012zd}
coupled with rapidly increasing computational recources open
the exciting possibility of testing chiral EFT in nuclear structure
and reaction calculations. A particularly promising new addition to
the existing \emph{ab initio} methods is provided by Nuclear Lattice
Effective Field Theory (NLEFT) \cite{Lee:2008fa}, an approach which naturally combines
chiral EFT with auxiliary-field quantum Monte Carlo lattice
simulations to access light- and medium-mass nuclei, see 
Refs.~\cite{Epelbaum:2011md,Epelbaum:2012qn,Epelbaum:2013paa,Elhatisari:2015iga}
for some most exciting results and the talk
by Mei{\ss}ner for a review \cite{UGMproc}.   

In this talk I will mainly focus on our recent work towards the development
of the new generation of NN potentials up to fifth order in the
chiral expansion \cite{Epelbaum:2014efa,Epelbaum:2014sza} along with the novel approach for error
analysis. I will also discuss selected applications beyond the NN
system. 
This paper is organized as follows. In section
\ref{sec2}, I will discuss the foundations of nuclear chiral EFT and 
focus on its application to the NN system. Section \ref{sec3} addresses
the issue of estimating the theoretical unsertainties due to the
truncation of the chiral expansion. Selected applications to the
three-nucleon system are presented in section \ref{sec4}. Finally, section
\ref{sec5} gives the outlook on ongoing and future research in this field.

\section{A new generation of chiral NN forces up to fifth order}
\label{sec2}

\subsection{Foundations}

The starting point in the
derivation of the nuclear forces and currents is the most general chiral
invariant effective Langangian for pions and nucleons.
Given the non-relativistic
nature of the problem at hand and the desire to treat nuclear
dynamics in the framework of the quantum mechanical $A$-body Schr\"odinger
equation, it is natural to utilize the heavy-baryon approach for the
description of the nucleon fields. For the explicit form of the $\pi N$
heavy-baryon Lagrangian up to fourth order, sufficient for
the applications discussed in this paper, see
Ref.~\cite{Fettes:2000gb}. The choice of pions and
nucleons as explicit degrees of freedom in the Lagrangian 
is dictated by the energy range the resulting EFT is
supposed to be applicable to. As an alternative, for few-nucleon reactions at very
low energies with typical three-momenta of the nucleons $| \vec p \, |$ being well
below the pion mass $M_\pi$ and/or for matching to lattice QCD results 
at unphysically heavy quark masses \cite{BeaneProc}, the so-called pion-less EFT
formulation is expected to provide a more efficient approach, see
the talk by Schindler for the application of this method to
parity-violating few-nucleon reactions \cite{SchindlerProc}. On the other hand, the explicit inclusion of the
$\Delta$(1232) resonance in the effective chiral Lagrangian along the lines
of Ref.~\cite{Hemmert:1997ye} is expected to improve the convergence pattern of
the chiral expansion and to increase its applicability range to higher
energies. Here and in
what follows, I will focus on the standard formulation of chiral
EFT based on pions and nucleons as the only degrees of freedom.

The effective chiral Lagrangian represents an infinite set of terms
which respect the approximate spontaneously broken chiral symmetry of
QCD. They can be classified by the number of
derivatives and/or pion mass insertions.  For applications in the
few-nucleon sector, it is convenient to assign the
chiral dimension $\Delta_i$ to a vertex $i$ which is defined as \cite{Weinberg:1990rz,Weinberg:1991um} 
\begin{equation}
\Delta_i = d_i + \frac{1}{2} n_i - 2,
\end{equation}
where $d_i$ is the number of derivatives or $M_\pi$-insertions while $n_i$ refers to the
number of nucleon field operators. Using dimensional counting, it
is straightforward to show that a 
connected diagram with $N$ nucleons and $L$ loops constructed out
of $V_i$ vertices of type $\Delta_i$ scales as $Q^\nu$ with the
chiral dimension $\nu$ given by 
\beq 
\label{WeinbergC}
\nu = - 4 + 2 N + 2 L + \sum_i V_i \Delta_i\,.
\eeq
Here,  $Q \in \{M_\pi / \Lambda_b, \; | \vec p \, |/\Lambda_b \}$ denotes the
expansion parameter of chiral EFT with $\Lambda_{b}$ being the breakdown scale
to be specified below. 
The above expression has a slightly different form than the one
derived originally  by Weinberg in  Ref.~\cite{Weinberg:1990rz} as
discussed  in Ref.~\cite{Epelbaum:2007us}. 
Notice
that since the effective Lagrangian involves only non-renormalizable interactions,
the  condition $\Delta_i \ge 0$ holds true for all $i$. This implies that $\nu$ is bounded from below
and that at each fixed order, a finite number of diagrams made out of a
finite number of vertices can contribute.    

Since Eq.~(\ref{WeinbergC}) is based solely on dimensional arguments,
it does not take into account the enhancement of reducible ladder-type
diagrams involving purely nucleonic intermediate states due to the
appearance of pinch singularities (in the static limit) when
performing integrals over zeroth components of the loop momenta. 
Clearly, such infrared singularities are regularized by the nucleon kinetic
energy term which has to be explicitly kept in the corresponding 
propagators. Still, the resulting contributions to the amplitude are
enhanced by powers of $m_N/| \vec p \, |$, where $m_N$ refers to the
nucleon mass, as compared to estimates based on 
dimensional analysis and underlying the derivation of
Eq.~(\ref{WeinbergC}). Fortunately, the contributions of the enhanced ladder-like diagrams
can be easily and efficiently resummed by solving the
LS integral equation (or its generalizations in the
case of three- and more-nucleon systems) whose kernel involves all
possible irreducible graphs which obey the scaling according to
Eq.~(\ref{WeinbergC}) and are derivable in perturbation
theory. This is the essence of what is commonly referred to as Weinberg's approach to
nuclear chiral EFT. The set of all possible irreducible
contributions to the scattering amplitude can be viewed as the
interaction part of the nuclear Hamiltonian and comprises two-, three-
and more-nucleon forces. The approach outlined above 
is straightforwardly generalizable to reactions involving external sources
and allows one to derive exchange currents consistent
with the nuclear forces.  

\begin{figure}[t]
\centering
\includegraphics[width=\textwidth,clip]{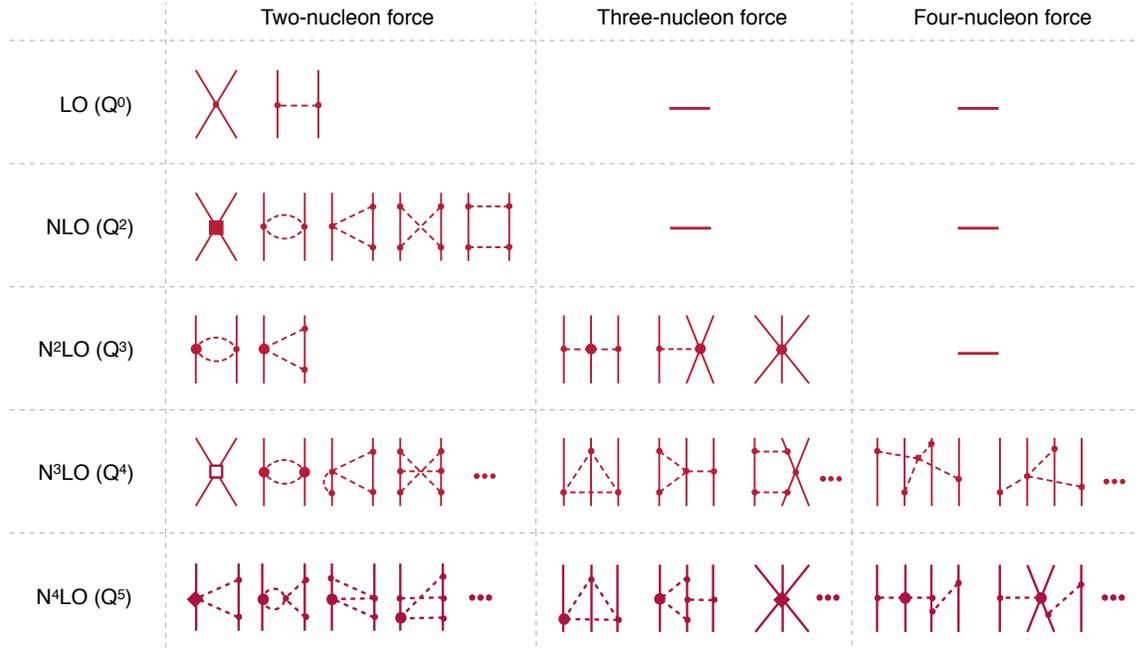}
\caption{
Chiral expansion of the nuclear forces. 
Solid and dashed lines refer to nucleons and pions,
  respectively. Solid dots, filled circles, filled rectangles, filled
  diamonds and open
  rectangles refer to the vertices of dimension 
$\Delta_i = 0$, $\Delta_i = 1$, $\Delta_i = 2$, $\Delta_i = 3$ and $\Delta_i =
4$, respectively. 
}
\label{fig1}       
\end{figure}

It is a simple exercise to enumerate the various diagrams which may 
contribute to the nuclear force at a given order $\nu$ by looking at
Feynman rules for the chiral Lagrangian
and applying Eq.~(\ref{WeinbergC}), see Fig.~\ref{fig1}.   
Here, it is understood that the shown diagrams only serve the purpose of
visualization of the corresponding contributions and do not have the 
meaning of Feynman graphs. In particular, one needs to separate out the
irreducible pieces in order to avoid double counting.  Notice further
that while one can draw three-nucleon diagrams at
next-to-leading order (NLO), the resulting contributions are either
reducible or suppressed by one power of $Q/m_N$ \cite{Weinberg:1992yk}.  
As an immediate consequence of the chiral power counting in
Eq.~(\ref{WeinbergC}), one observes the suppression of many-body
forces \cite{vanKolck:1994yi}, the feature, that has always been assumed 
but could be justified only in the context of chiral EFT. 

So far, I have not yet discussed the important issue of identifying 
the irreducible parts of the scattering amplitude, i.e.~the actual
derivation of the nuclear forces. This can be achieved using a
veriety of
approaches. In particular, old-fashioned time-ordered perturbation
theory was employed by Weinberg in his original papers 
\cite{Weinberg:1991um,Weinberg:1992yk} and by
Ordonez {\it et al.} in their first numerical analysis of NN
scattering \cite{Ordonez:1995rz}. More recently, this method was
taken over by the Jlab-Pisa group to derive electromagnetic \cite{Pastore:2009is,Pastore:2011ip,Piarulli:2012bn} and weak
\cite{Baroni:2015uza} exchange currents, see the talk by Schiavilla \cite{SchiProc} for more
details. We used another well-known technique 
commonly referred to as the method of unitary transformation (UT) \cite{TMO,Okubo:1954zz}. 
The main idea of this approach consists in block-diagonalization of the
pion-nucleon Hamiltonian in Fock space via a suitable chosen UT.
The corresponding unitary operator can be determined
perturbatively employing the chiral power counting as explained in
detail in
Refs.~\cite{Epelbaum:1998ka,Epelbaum:2002gb,Epelbaum:2005pn,Epelbaum:2007us}, see also
Ref.~\cite{Krebs:2004st} for a closely related approach. The method of
UT can be straightforwardly applied to derive
the exchange currents \cite{Kolling:2009iq,Kolling:2011mt}. Finally, 
nuclear forces can also be defined by calculating Feynman diagrams and
matching
the results to the LS equation, see Ref.~\cite{Kaiser:1997mw}
for more detail. 

Irregardless of the employed  approach to derive nuclear potentials, it is
important to keep in mind that, in contrast to S-matrix elements,
nuclear forces and exchange currents are not uniquely defined
and can always be changed by unitary transformations 
or, equivalently, by changing the basis in the Fock space. The intrinsic
unitary ambiguity of the Hamiltonian and exchange currents turns out
to play the crucial role in maintaining their renormalizability. More
precisely, these quantities will generally \emph{not} stay finite
after evaluating the loop integrals and replacing the bare LECs with
the renormalized ones. This is in a strong contrast with
on-shell scattering amplitudes calculated in ChPT, for which all
ultraviolet divergences do, of course, cancel against the corresponding
counterterms. This issue was first raised in Ref.~\cite{Epelbaum:2006eu}, where
renormalizability of the Hamiltonian could finally be established
for a specific choice of the
unitary transformations (i.e. for a particular choice of basis in the Fock
space). We also followed the same approach of systematically
exploiting the freedom in the choice of 
the basis in the Fock space to maintain
renormalizability of the electromagnetic exchange charge and
current operators to leading one-loop order in Refs.~\cite{Kolling:2009iq,Kolling:2011mt}. This is in contrast with the
results of the Jlab-Pisa group \cite{Pastore:2009is,Pastore:2011ip,Piarulli:2012bn,Baroni:2015uza}, whose expressions for the
one-pion exchange contributions contain ultraviolet divergences  
which are not cancelled by counterterms from the effective
chiral Lagrangian. 

Having outlined the various techniques to derive the nuclear
Hamiltonian, I now briefly discuss the current state of the art
for the contributions shown in Fig.~\ref{fig1}. The NN force up to
next-to-next-to-leading order (N$^2$LO) has been known since about two
decades and was derived using all three approaches mentioned above 
\cite{Ordonez:1995rz,Kaiser:1997mw,Epelbaum:1998ka}. The order-$Q^4$
(N$^3$LO) corrections to the NN potential 
have been worked out in a series of papers by Kaiser
\cite{Kaiser:1999ff,Kaiser:1999jg,Kaiser:2001pc,Kaiser:2001at} about  fifteen
years ago using an S-matrix based approach and employing Cutkosky's
cutting rules to simplify the calculation of loop integrals. 
These results provided a basis for the development of the first generation of
chiral NN potentials at N$^3$LO \cite{Entem:2003ft,Epelbaum:2004fk}. 
Very recently, the same technique was used by Entem {\it et al.}~\cite{Entem:2014msa}
to work out the order-$Q^5$ (N$^4$LO) corrections to the two-pion
exchange NN potential\footnote{We have independently calculated these
  contributions \cite{Epelbaum:2014sza} and have verified the results
  obtained 
  by Entem {\rm et al.}.}  
and even the dominant order-$Q^6$ (N$^5$LO)
terms \cite{Entem:2015xwa}. 
Notice that the leading $3\pi$-exchange potentials at N$^3$LO
turn out to be negligibly weak \cite{Kaiser:1999ff,Kaiser:1999jg} and were not included explicitly in Ref.~\cite{Epelbaum:2004fk}. 
While the subleading $3\pi$-exchange potentials at N$^4$LO are
stronger in magnitude, their short-range nature permits their implicit
representation via contact interactions \cite{Epelbaum:2014sza}.

The expressions for the leading 3NF at N$^2$LO resulting from 
tree-level diagrams shown in Fig.~\ref{fig1} have been known since
a long time \cite{vanKolck:1994yi,Epelbaum:2002vt} and depend on two
LECs $c_D$ and $c_E$ which accompany the contact interactions in the
second and third diagram, respectively. The first corrections at N$^3$LO 
encompass numerous one-loop diagrams. Their calculation has
been accomplished in Refs.~\cite{Ishikawa:2007zz,Bernard:2007sp,Bernard:2011zr}. 
Interestingly, the resulting contributions do not involve any unknown
LECs. The long- and intermediate-range corrections to the 3NF at
N$^4$LO have been derived recently using the method of unitary
transformation \cite{Krebs:2012yv,Krebs:2013kha} while the purely
contact 3NF terms have been worked out in
Ref.~\cite{Girlanda:2011fh} and involve 10 unknown LECs. 
The derivation of the remaining contributions involving NN contact interactions is in
progress \cite{KrebsProc}. Finally, the leading four-nucleon force at N$^3$LO has been calculated
in Refs.~\cite{Epelbaum:2006eu,Epelbaum:2007us} and is also parameter-free. The
corrections at N$^4$LO  have not been studied yet.

\subsection{Renormalization of the
Schr\"odinger equation and regularization of the potential}
\label{SubSecRen}

The chiral nuclear potentials discussed in the previous section are
constructed to be used in the $A$-body Schr\"odinger equation 
\beq
\bigg[ \bigg( \sum_{i=1}^A \frac{- \Delta_i}{2 m_N} + \mathcal{O}
\big( m_N^{-3} \big) \bigg) + V_{2N} + V_{3N} + V_{4N} + \ldots \bigg]
| \Psi \rangle = E | \Psi \rangle\,,
\eeq
where the nuclear forces $V_{2N}$, $V_{3N}$, $V_{4N}$, $\ldots$, are
derived by means of the chiral expansion as discussed above. Consider,
for the sake of simplicity, the NN system where the Schr\"odinger
equation can be conveniently cast into the LS
integral equation for the $T$-matrix. In the operator form, this
equation reads 
\beq
\hat T = \hat V_{NN} + \hat V_{NN} \hat G_0 \hat T = \hat V_{NN} +
\hat V_{NN} \hat G_0 \hat V_{NN} + \hat V_{NN} \hat G_0 \hat V_{NN} \hat
G_0 \hat V_{NN} + \ldots\,,
\eeq
where $\hat G_0$ is the free resolvent operator. Iterations of the NN
potential in the LS equation generate ultraviolet (UV) divergences which
are cancelled by counterterms of the contact interaction type. While
this cancellation certainly holds true for the potential calculated up to an
infinite order in the chiral expansion which involves all
possible counterterms compatible with the symmetries, it is not ensured anymore if a truncated
expression for $V_{NN}$ is employed.  Indeed, it is easy 
to verify that the LS equation for the leading-order (LO) one-pion
exchange potential (OPEP) is linearly divergent. Consequently, its iterations
in any given spin-triplet partial wave generate UV divergences whose 
cancellation requires the introduction of infinitely many
counterterms with increasing powers of momenta and/or $M_\pi$. 

The simplest way to circumvent this problem is by treating the OPEP
perturbatively \cite{Kaplan:1998we}. The resulting EFT
approach, however, fails (at least) in certain spin-triplet channels
due to the lack of convergence \cite{Fleming:1999ee}. 
An alternative solution has been proposed recently in Ref.~\cite{Epelbaum:2012ua}, where
it was pointed out that the linearly divergent UV behavior of the LS
equation for the OPEP can be traced back to 
the non-relativistic
expansion being carried out prior to solving the dynamical equation. 
Using, instead of the LS equation, a three-dimensional integral
equation which maintains relativistic elastic unitarity such as
e.g.~the Kadyshevsky equation \cite{Kadyshevsky:1967rs} results in a milder UV
divergence of a logarithmic type. The resulting LO equation for the
OPEP and derivative-less contact interactions is renormalizable in the
sense that all UV divergences generated by iterations can be
absorbed into the two LO derivative-less contact interactions. Clearly, the intrinsically
non-relativistic nature of the problem is not affected by this
procedure, and it is perfectly fine to perform, if desired, a
$1/m_N$-expansion of the amplitude after solving the integral equation
\cite{Epelbaum:2012ua}. This renormalizable approach outlined above 
permits a complete removal of the UV cutuff and
has been applied at LO to chiral extrapolations \cite{Epelbaum:2013ij} and to the
electromagnetic form factors of the deuteron
\cite{Epelbaum:2013naa}. Higher-order corrections have to 
be included perturbatively in order to maintain 
renormalizability.\footnote{Higher-order contact interactions can also
  be
 treated non-perturbatively, see Refs.~\cite{Epelbaum:2015sha}
for more details.}  

While the approach outlined above is the only way to completely remove
the UV cutoff in calculations with the OPEP being treated non-perturbatively 
within the framework of an EFT I am aware of\footnote{Several authors have explored the possibility
  of removing the UV regulator in the context of the LS equation
  via performing self-adjoint extentions of singular
  potentials due to one- (and more-) pion exchanges, see
  e.g.~\cite{Nogga:2005hy,PavonValderrama:2005wv,Beane:2001bc}. Such manifestly
  non-perturbative approaches have been demonstrated in Ref.~\cite{Epelbaum:2009sd}
  to be, in general, incompatible with the principles of EFT in spite
  of being able to yield finite results for the scattering
  amplitude. It is important to remember that the UV regulator can
  be
  removed from the calculations only after \emph{all} UV
  divergences have been subtracted.}, its extention to
higher orders and/or  heavier systems represents a challenging task. 
An alternative and much simpler (but still valid from the EFT point of view)
approach is to introduce a finite UV regulator which is to be chosen of the
order of the relevant hard scale in the problem as advocated in
Refs.~\cite{Lepage:1997cs,Epelbaum:2006pt}. 
As already mentioned, taking the values of the
UV cutoff $\Lambda$ in the LS equation well beyond the pertinent hard scale 
may result in the breakdown of low-energy theorems \cite{Epelbaum:2009sd} which
signals the violation of basic principles of EFT. I refer interested
readers to our work \cite{Epelbaum:2009sd}, which provides an explicit
example of such a failure and brings further analytical insights into
these topics by considering an exactly solvable model. The
finite-regulator scheme is very well suited for applications using \emph{ab initio} few- and many-body
methods and has been adopted in the most sophisticated nuclear chiral EFT
calculations carried out up to date. This is the approach I will focus
on in the following. 

For calculations with a finite UV cutoff, it is important to
employ such a functional form of the regulator which avoids
introducing unnecessary artifacts. In the new generation of chiral
potentials of Refs.~\cite{Epelbaum:2014efa,Epelbaum:2014sza}, we used
a coordinate-space regularization procedure for the 
long-range components $V_{\rm long}  (\vec r \, )$,  
\begin{equation}
\label{RegCoord}
V_{\rm long} (\vec r \, ) \to V_{\rm long}^{\rm Reg.} (\vec r \, ) =
V_{\rm long} (\vec r \, ) \; f\bigg( \frac{r}{R} \bigg) 
\quad 
\mbox{with}
\quad
f(x) = \Big(1 - \exp(-x^2) \Big)^6 \,,
\end{equation}
where the cutoff distance $R$ is chosen in the range of $R=0.8\ldots
1.2\,$fm in agreement with the expected breakdown distance of $\sim
0.8\,$fm of the
chiral expansion for the pion-exchange potential \cite{Baru:2012iv}. 
A similar regulator was employed in
Refs.~\cite{Gezerlis:2013ipa,Piarulli:2014bda}. In momentum space, 
the regularization takes the form 
\beq
V_{\rm long} ( \vec q \, )  \to V_{\rm long}^{\rm reg} ( \vec q \, ) = V_{\rm long} ( \vec q \, )  -
\int \frac{d^3 l}{(2 \pi)^3} V_{\rm long} (\vec l \, ) \, {\rm FT}_{\vec q - \vec l}  \left[ 1 - f
\right] \,,
\eeq
where ${\rm FT}$ stays for the Fourier-Transform and $\vec q$ is the
momentum transfer.  Given that ${\rm
  FT} \left[ 1 - f \right]$ is a short-range operator, it is clear
that the regulator does, per construction, not affect the long-range
part of the interaction and thus maintains the analytic structure of
the amplitude in the low-energy domain. This feature is in contrast
with the non-local momentum-space regulator employed in the
first-generation NN potentials of Refs.~\cite{Entem:2003ft,Epelbaum:2004fk} of the type 
\beq
\label{RegOld}
V(\vec p, \, \vec p\,') \to V^{\rm reg}  (\vec p, \, \vec p\,') =
V(\vec p, \, \vec p\,') \exp \bigg( -\frac{p^{2n} + p
  '^{2n}}{\Lambda^{2n}} \bigg),  \quad n =2,3\,,
\eeq
where $\vec p$, $\vec p \, '$ are the initial and final momenta of the
nucleons in the center of mass system (CMS), which distorts the
long-range part of the interaction. 
Another
advantage of the regulator in Eq.~(\ref{RegCoord}) is that it cuts off precisely the undesired
\emph{short-range} components of the pion exchange contributions
which cannot be meaningfully predicted in chiral EFT instead of  
their \emph{large-momentum} parts as does the non-local
regulator in Eq.~(\ref{RegOld}). 
This makes the additional spectral-function regularization (SFR) \cite{Epelbaum:2003gr}
of the two-pion exchange components, which was used e.g.~in Refs.~\cite{Epelbaum:2004fk,Marji:2013uia} to tame the unphysically 
strong attraction at short distances at N$^2$LO \cite{Kaiser:1997mw}, obsolete. 
This is a particularly welcome feature in view of the
ongoing and upcoming 3NF studies, in which  the implementation of the
SFR would be rather non-trivial. The insensitivity of the calculated
NN observables to the value of the exponent in Eq.~(\ref{RegCoord}) is
demonstrated in \cite{Epelbaum:2014efa}. 
For contact interactions, we used in 
Refs.~\cite{Epelbaum:2014efa,Epelbaum:2014sza} a non-local
Gaussian regulator in momentum space with the cutoff set to $\Lambda = 2/R$.

\subsection{Determination of the LECs}

\begin{figure}[t]
\centering
\includegraphics[width=0.8\textwidth,clip]{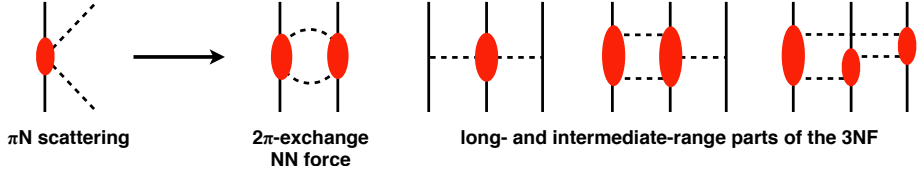}
\caption{The long-range part of the nuclear force is completely
  predicted by the chiral symmetry of QCD and experimental information
  on the pion-nucleon system. }
\label{fig2}       
\end{figure}

I am now in the position to specify the employed values of the various
LECs and begin with the long-range part of the potential due to
exchange of pion(s). Here, the framework of chiral EFT shows
its full power by allowing one to \emph{predict} the long-range
part of the nuclear force in a parameter-free way using the available
experimental information on the pion-nucleon system and exploiting
the constraints due to the chiral symmetry of QCD as visualized
schematically in Fig.~\ref{fig2}. 
At orders N$^2$LO, N$^3$LO and N$^4$LO, one needs to specify the values
of the order-$Q^2$, order-$Q^3$ and order-$Q^4$ $\pi N$ LECs $c_i$,
$d_{i}$ and $e_i$, respectively.  At N$^2$LO and N$^3$LO, we used in~\cite{Epelbaum:2014efa} the
values of $c_1 = -0.81$, $c_2 = 3.28$, $c_3=-4.69$, $c_4=3.40$, 
$\bar d_1 + \bar d_2 = 3.06$, $\bar d_3 = -3.27$, $\bar d_5 = 0.45$
and $\bar d_{14} - \bar d_{15} = -5.65$ from the order-$Q^3$
fits to $\pi N$ data in the physical region \cite{Fettes:1998ud} and inside the
Mandelstam triangle \cite{Buettiker:1999ap}. Further, the LEC $d_{18}$ is
adjusted to reproduce the observed value of the Goldberger-Treiman
discrepancy. Here and in the following, the values of
the LECs are given in  units of GeV$^{-n}$. The bars over
the LECs indicate that I am  using the convention of Ref.~\cite{Fettes:1998ud} by
setting the dimensional regularization scale equal to the pion mass. At N$^4$LO,
we employ the values from our order-$Q^4$ fit to Karlsruhe-Helsinki  
partial-wave analysis of $\pi N$ scattering \cite{Krebs:2012yv}, namely:
$c_1 = -0.75$, $c_2 = 3.49$, $c_3=-4.77$, $c_4=3.34$, 
$\bar d_1 + \bar d_2 = 6.21$, $\bar d_3 = -6.83$, $\bar d_5 = 0.78$,
$\bar d_{14} - \bar d_{15} = -12.02$, $\bar e_{14}= 1.52$ and $\bar
e_{17} = -0.37$. These values are in a reasonable agreement with the
ones of Ref.~\cite{Wendt:2014lja} from a $Q^4$ fit to $\pi N$
scattering data  and with the ones of Ref.~\cite{Hoferichter:2015tha} from matching to the
subthreshold coefficients obtained from the Roy-Steiner analysis of $\pi N$
scattering \cite{Hoferichter:2012wf}. Still, the differences between the values
found in these determinations indicate sizable uncertainties in the LECs, 
whose impact on the nuclear forces needs to be addressed in future studies. 

I now turn to the short-range part of the potential. Adopting  the
standard power counting rules for short-range operators, which are
based on naive dimensional analysis\footnote{For alternative
  suggestions see e.g.~Refs.~\cite{Nogga:2005hy,Long:2011xw,Birse:2010fj}.}, one has to take into account
(in the isospin limit) $2$ order-$Q^0$ contact terms at LO, $7$ additional order-$Q^2$
contact interactions at NLO and N$^2$LO and  $15$ additional order-$Q^4$
contact terms at N$^3$LO and N$^4$LO. This leaves one with 24 unknown LECs in
total, which have been determined from a fit to np
and pp S-, P- and D-waves and to the mixing angles
$\epsilon_1$ and $\epsilon_2$ of the Nijmegen partial wave analysis
(NPWA) \cite{Stoks:1993tb} for five different choices of the
regulator $R$ 
\beq
\label{RegSp}
\Big\{R_1, \, R_2, \, R_3, \, R_4, \, R_5 \Big\} =  \Big\{
0.8\mbox{ fm}, \;  0.9\mbox{ fm}, \;  1.0\mbox{ fm}, \;  1.1\mbox{ fm}, \;  1.2\mbox{ fm}\Big\}\,.
\eeq
We also included  2 (3)
isospin-violating contact interactions up to N$^3$LO (at N$^4$LO) which
account for isospin-breaking effects in the $^1$S$_0$ partial wave. 
A detailed description of the fit procedure and the treatment of relativistic and
isospin-breaking terms can be found in the original papers
\cite{Epelbaum:2014efa,Epelbaum:2014sza}. Here, we only emphasize that
all LECs were found to be of a natural size for all five choices of the
cutoff $R$, see Table II of Ref.~\cite{Epelbaum:2014efa}.\footnote{The
LECs accompanying the order-$Q^4$ S-wave contact interactions, however, turn out
to be significantly larger in magnitude than the other LECs.}

\subsection{Results for phase shifts}
\label{PhS}

\begin{figure}[t]
\centering
\includegraphics[width=\textwidth,clip]{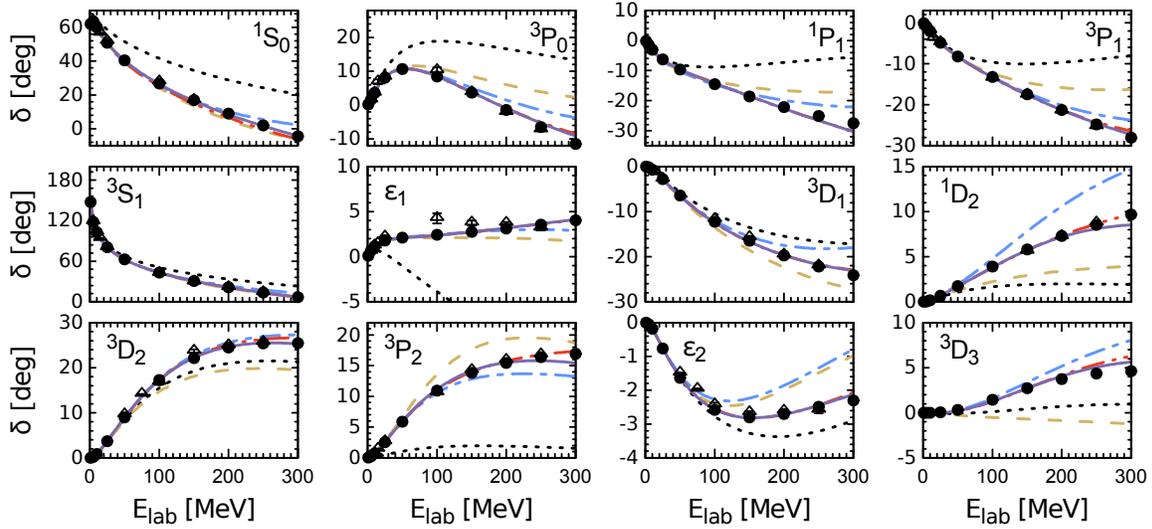}
\caption{Chiral expansion of the np phase shifts for the
  cutoff $R=0.9\,$fm in
  comparison with NPWA  \cite{Stoks:1993tb}  (solid dots) and the GWU single-energy np partial wave
  analysis \cite{Arndt:1994br}  (open triangles). Dotted (black), dashed (brown),
  dash-dotted (blue), dash-double-dotted (red) and solid (violet)
  lines show the results at LO, NLO, N$^2$LO, N$^3$LO and N$^4$LO,
  respectively.  }
\label{fig3} 
\end{figure}

The resulting np phase shifts are plotted in
Fig.~\ref{fig3} as functions of the laboratory energies in comparison
with the NPWA and the single-energy partial wave analysis of
Ref.~\cite{Arndt:1994br} for the cutoff choice of $R=0.9\;$fm. 
In all cases, one observes a good and natural convergence
pattern with the results at N$^4$LO being almost indistinguishable
from those at N$^3$LO. 

To get more quantitative insights, it is instructive to look at $\chi^2$ per datum for the
description of the Nijmegen np and pp phase shifts. 
For a phase shift or mixing angle $\delta$ in a channel $X$ at a given
energy, we assign the error of the NPWA to be 
\beq
\Delta_X = \max \Big( \Delta^{\rm NPWA}_X, \; | \delta_X^{\rm NijmI} -
  \delta_X^{\rm NPWA} |,\;
| \delta_X^{\rm NijmII} -
  \delta_X^{\rm NPWA} |, \; | \delta_X^{\rm Reid93} -
  \delta_X^{\rm NPWA} | \Big)\,,
\eeq
where $\Delta^{\rm NPWA}_X$ is the statistical error of the NPWA \cite{Stoks:1993tb},  while 
$\delta_X^{\rm
  NijmI}$,  $\delta_X^{\rm NijmI}$ and  $\delta_X^{\rm Reid93}$ refer
to the results based on the Nijmegen I, II and Reid93 NN potentials of
Ref.~\cite{Stoks:1994wp} which provide a nearly optimal description of
the same database as employed in the NPWA and can be regarded as alternative PWA. 
I emphasize, however, that the resulting $\chi^2$ per datum
does not allow for a simple statistical interpretation due to the
artificially chosen errors and, in particular, can take values $< 1$ without
indicating any inconsistencies. Therefore, in order to avoid a possible confusion
or misinterpretation, I will use the notation $\tilde \chi^2$.  

In table \ref{tab_chi2_2}, I give $\tilde \chi^2$ per datum for the description
of the np and pp phase shifts of the NPWA at the energies of 
$E_{\rm lab}=$1, 5, 10, 25, 50, 100, 150 and 200 MeV. 
\begin{table}[t]
\centering
\caption{$\tilde \chi^2/{\rm datum}$ for the description of the Nijmegen
np and pp phase shifts \cite{Stoks:1993tb} 
for the cutoff $R=0.9$ fm. The numbers
in the round brackets give the number of adjustable isospin-invariant
contact interactions at the corresponding order with the subscripts
referring to the number of isospin-breaking contact terms.  Only those
channels are included which have been used in the N$^3$LO/N$^4$LO fits,
namely the S-, P- and D-waves and the mixing angles $\epsilon_1$ and
$\epsilon_2$.
\label{tab_chi2_2}}
\smallskip
\begin{tabularx}{\textwidth}{@{\extracolsep{\fill}}cccccc}
\hline
\noalign{\smallskip}
 $E_{\rm lab}$ bin &  LO ($2_{[2]}$) &  NLO ($+\, 7_{[0]}$)  &
                                                               N$^2$LO
                                                               ($+\,
                                                               0_{[0]}$)
  &  N$^3$LO ($+\, 15_{[0]}$)  &  N$^4$LO ($+\, 0_{[1]}$) 
\smallskip
 \\
\hline 
\multicolumn{5}{l}{neutron-proton phase shifts} \\ 
0--100 & 360 & 31 & 4.5 & 0.7  & 0.3\\ 
0--200 & 480 & 63 & 21 & 0.7  & 0.3\\ [4pt]
\hline 
\multicolumn{5}{l}{proton-proton phase shifts} \\ 
0--100 & 5750 & 102 & 15 & 0.8 & 0.3 \\ 
0--200 & 9150 & 560 & 130 & 0.7 & 0.6 \\[3pt]
\hline 
\end{tabularx}
\end{table}
As expected, one observes the improved description of the phase
shifts with increasing chiral order. It is
particularly encouraging to see a strong reduction in the $\tilde \chi^2$
when going from NLO ($Q^2$) to N$^2$LO ($Q^3$) and from N$^3$LO ($Q^4$) to
N$^4$LO ($Q^5$), which originates entirely from the corresponding two-pion
exchange (TPE) components without invoking new
parameters.\footnote{Except for the  np $^1$S$_0$ partial wave, where
one additional isospin-breaking contact interaction is included at N$^4$LO.} These results constitute an
important consistency check of the theoretical approach and provide a
beautiful illustration of its predictive power. In particular, one observes
a strong decrease in the value of $\tilde \chi^2$ per datum at
N$^2$LO which is solely due to inclusion of the parameter-free
order-$Q^3$ TPE. Similarly, we find  a
significant decrease in $\tilde \chi^2$ per datum at N$^4$LO which,
for the pp case, again emerges entirely from the predicted
parameter-free order-$Q^5$ contributions to the TPE 
potential.  
The obtained 
results suggest --  fully in line with the
Weinberg power counting \cite{Weinberg:1990rz} -- 
that the theoretical uncertainty at NLO
and N$^3$LO is dominated by the neglected TPE 
contributions at orders $Q^3$ and $Q^5$, respectively. Indeed, if certain
order-$Q^4$ and order-$Q^6$ contact interactions would have to be promoted
to lower orders in violation with naive dimensional
analysis as suggested e.g.~in
\cite{Nogga:2005hy,Birse:2010fj,Long:2011xw}, the inclusion of the
order-$Q^3$ and order-$Q^5$ TPE
contributions alone would not result in the
improved accuracy of the fits at N$^2$LO and N$^4$LO. 

\begin{figure}[t]
\centering
\includegraphics[width=0.9\textwidth,clip]{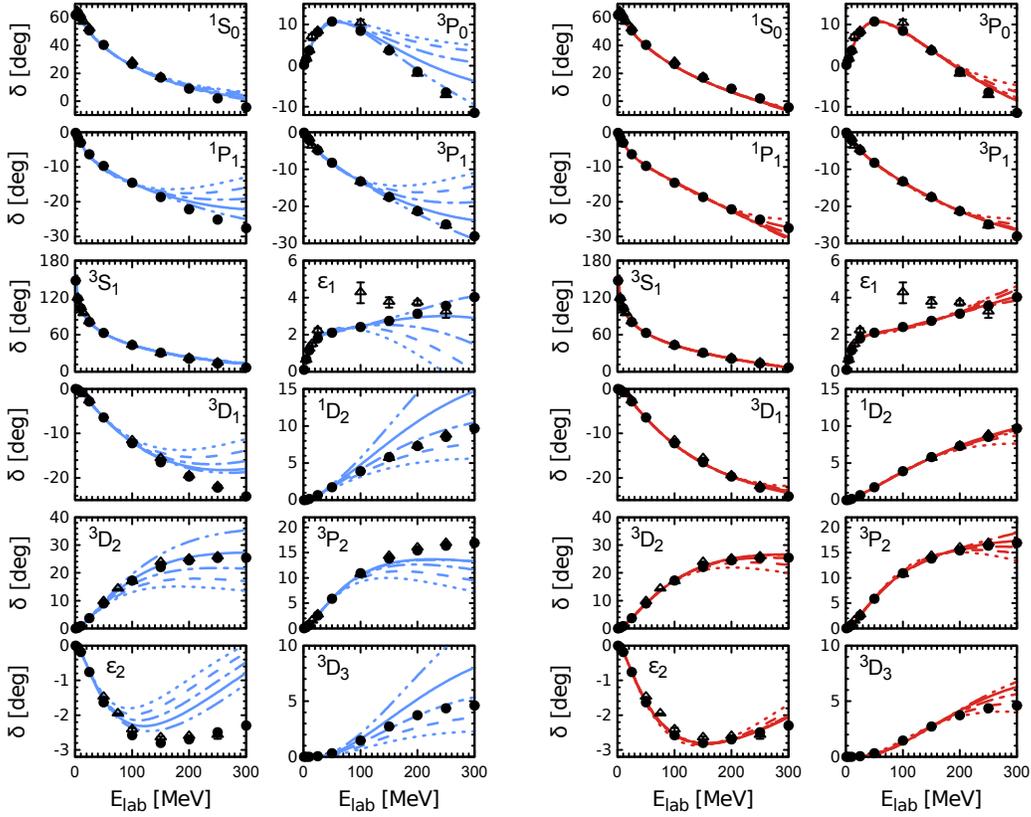}
\caption{Cutoff dependence of the phase shifts calculated at N$^2$LO (left panel) and N$^3$LO (right panel). 
Dashed-double-dotted, solid, dashed-dotted, dashed and dotted lines
show the results for $R=R_1, \ldots , R_5$ as defined in Eq.~(\protect\ref{RegSp}),
respectively. For remaining notation see Fig.~\protect\ref{fig3}.
\label{fig:phases_cutoff} }
\label{fig4}       
\end{figure}

I now briefly address the residual cutoff dependence of our
results. Fig.~\ref{fig4} shows the np phase shifts at N$^2$LO and
N$^3$LO for all considered choices of the regulator.
 As expected, the residual cutoff dependence at N$^2$LO is efficiently
absorbed into redefinition of the order-$Q^4$ contact
interactions at N$^3$LO. I do not show the results at N$^4$LO, but they turn out to
be very similar to those at N$^3$LO what
concerns the dependence on the regulator $R$.  

It is also instructive to look at $\tilde \chi^2$ per datum for the
reproduction of the phase shifts of the NPWA as a function
of the cutoff $R$. Here, for the sake of brevity, I restrict
myself to N$^3$LO and to the single energy bin of $E_{\rm lab} = 0-200$ MeV. We find
the following pattern for np phase shifts by decreasing the values of
the regulator starting from the softest choice of $R=1.2\;$fm:
\begin{displaymath}
\tilde \chi^2/\mbox{datum} \; = \; 1.8_{R=1.2\; \mbox{\small fm}} \;
\to \; 
0.8_{R=1.1\; \mbox{\small fm}} \;
\to \; 
0.6_{R=1.0\; \mbox{\small fm}} \;
\to \; 
0.7_{R=0.9\; \mbox{\small fm}} \;
\to \; 
0.8_{R=0.8\; \mbox{\small fm}} \,,
\end{displaymath}
while the results for pp channels are: 
\begin{displaymath}
\tilde \chi^2/\mbox{datum} \; = \; 8.2_{R=1.2\; \mbox{\small fm}} \;
\to \; 
2.2_{R=1.1\; \mbox{\small fm}} \;
\to \; 
0.6_{R=1.0\; \mbox{\small fm}} \;
\to \; 
0.7_{R=0.9\; \mbox{\small fm}} \;
\to \; 
2.1_{R=0.8\; \mbox{\small fm}} \,.
\end{displaymath}
As expected, the softest choice of the regulator of
$R=1.2\;$fm leads to the worst quality of the fit indicative of the 
largest amount of cutoff artifacts. 
Decreasing the value of $R$ improves the
description of phase shifts. The improvement stops around $R=0.9-1.0\;$fm, and the fits
start deteriorating for $R=0.8\;$fm. This is exactly the pattern one
expects to observe in calculations within this theoretical framework
as discussed in section \ref{SubSecRen} and in
Refs.~\cite{Lepage:1997cs,Epelbaum:2006pt,Epelbaum:2009sd}.

\section{Theoretical uncertainty}
\label{sec3}

Uncertainty quantification is a key for performing a meaningful
comparison between theoretical predictions/postdictions and
experimental data \cite{Furnstahl:2014xsa}.
The various sources of uncertainty in our calculations 
include:  (i) uncertainty in the knowledge of $\pi N$
LECs which determine the long-range part of the interaction, (ii)
systematic and statistical uncertainty in the determination of NN
contact interactions, (iii) uncertainty in the NPWA used to determine the LECs and (iv) 
systematic uncertainty due to truncation of the chiral expansion. The impact of the statistical uncertainty of the LECs accompanying
contact interactions on np and pp phase shifts was quantified in
Ref.~\cite{Ekstrom:2014dxa} at N$^2$LO and found to be negligible at the accuracy
level of our calculations. Similarly, we believe that the systematic uncertainty
of the NPWA used as input in our analysis has a minor effect on our
results, but this needs to be explicitly verified by using real
data in the fits. Here I will focus on quantifying the
effects of the truncated higher-order terms in the chiral expansion
which I expect to be the dominant source of uncertainty in our
calculations. 

Residual cutoff dependence of observables provides one possible way 
to estimate theoretical uncertainty due to truncation of the chiral
expansion. However, as pointed out already in \cite{Epelbaum:2004fk}, such an approach
suffers from several drawbacks: First, the residual cutoff dependence
 measures the impact of neglected contact interactions
which contribute only at \emph{even} orders $Q^{2n}$, $n=0,1,2,
\ldots$. This results in underestimation of uncertainties at NLO and N$^3$LO. 
Secondly, the available cutoff range is, in practice, rather
limited, and attempts to increase the cutoff range by employing softer regulators 
are likely to cause an unnecessary increase of finite-cutoff artifacts. 
The ``cutoff bands'' can, therefore, generally not be expected to
provide a reliable  estimation of the theoretical uncertainty, see
\cite{Epelbaum:2004fk,Epelbaum:2014efa} for more detail. 

\begin{figure}[t]
\centering
\includegraphics[width=0.6\textwidth,clip]{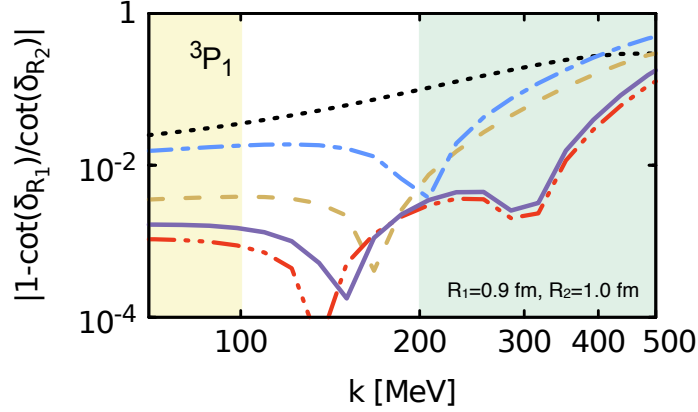}
\caption{Error plot for np $^3$P$_1$ phase shift. Black dotted, brown
  dashed, blue dashed-dotted, red dashed-double-dotted and violet
  solid lines show the results at LO, NLO, N$^2$LO, N$^3$LO and
  N$^4$LO, respectively. The light-shaded yellow (green) area shows
  the range of momenta where the error is expected to be dominated by
  neglected $M_\pi/\Lambda_b$-contributions
  ($p/\Lambda_b$-contributions).
}
\label{fig5}       
\end{figure}

In order to circumvent these problems, we proposed in Ref.~\cite{Epelbaum:2014efa} a novel approach to
error analysis by directly estimating the size of neglected
higher-order terms. Let $X(p)$ be a given observable with $p$
referring to the corresponding momentum scale and $X^{(i)} (p)$, $i = 0, 2, 3, \ldots$, 
a prediction at order $Q^i$ in the chiral expansion. 
We  further define the order-$Q^i$ corrections to $X(p)$ as 
\begin{equation}
\Delta X^{(2)} \equiv X^{(2)} - X^{(0)}, \quad \mbox{and}\quad
\Delta X^{(i)} \equiv X^{(i)} - X^{(i-1)}\;\;\mbox{for} \; \; i \geq 3 \,,
\end{equation}  
so that the chiral expansion for $X$ up to order $Q^i$ takes the form  
\begin{equation}
\label{Expan}
X^{(i)} = X^{(0)}+ \Delta X^{(2)}+ \ldots +  \Delta X^{(i)}, \quad
\mbox{with}\quad 
X^{(i)}= \mathcal{O} ( Q^i X^{(0)})\,.
\end{equation}  
One can now use the information on $X^{(0)}$ and $\Delta
X^{(j)}$, $j \le i$, available upon
performing explicit calculations, to quantify 
the theoretical uncertainty $\delta X^{(i)}$ of the order-$Q^i$ result
by estimating the size of neglected higher-order terms. Specifically, the
following procedure was employed in Refs.~\cite{Epelbaum:2014efa,Epelbaum:2014sza}:
\begin{equation}
\label{Err}
\delta X^{(0)} = Q^2 | X^{(0)} |,  \quad \quad
\delta X^{(i)} = \max\limits_{2\le j \le i} \Big( Q^{i+1} | X^{(0)}
                   |, \,  Q^{i+1-j} | \Delta X^{(j)} | \Big) \;\;\mbox{with}\;\;  i
                   \geq 2,   
\end{equation}
where the expansion parameter $Q$ was chosen as  $Q = \max
(p/\Lambda_b, M_\pi/\Lambda_b )$, subject to the additional constraint 
\begin{equation}
\label{ErrExplicit}
\delta X^{(i)} \,  \ge \,   \max\limits_{j,k} \Big( \big| X^{(j \ge i)} -  X^{(k \ge i)} \big| \Big).
\end{equation}
The breakdown scale $\Lambda_b$ of the chiral expansion can be
estimated e.g.~from error plots such as the ones proposed by Lepage \cite{Lepage:1997cs}
or by Grie{\ss}hammer, see \cite{GrieProc} for more detail. An example
of the latter one for the $^3$P$_1$ np partial wave is shown in Fig.~\ref{fig5}.  
The depicted error function $|  1 - \cot \delta_{R_1}(k) / \cot
\delta_{R_2} (k) |$ with $k$ referring to the CMS momentum measures
the resudual cutoff dependence of the phase shift. The observed flat
behavior of the error function for $k$ well below $M_\pi$
(yellow-shaded area) is in line with
the expectation that the error in this regime is dominated by
neglected $M_\pi/\Lambda_b$-terms. On the other hand, the
uncertainty at large momenta (green-shaded area) is dominated 
by neglected $k/\Lambda_b$-terms, and one indeed observes an increased
slope at N$^3$LO/N$^4$LO versus NLO/N$^2$LO versus
LO.\footnote{Interpretation of such error plots should,
  however, be taken with care. In particular, the dependence of the
  lines in a
  double-logarithmic plot deviates from the linear one in the presence
  of the long-range interaction due to exchange of pion(s). } 
The observed nearly identical slopes at NLO and N$^2$LO as well as N$^3$LO and N$^4$LO reflect 
the same number of contact interactions at those orders. The spikes in
the plot emerge when $\delta_{R_1} = \delta_{R_2}$ for some value
of $k$ and should be ignored. The error plots can be used to read off 
the breakdown scale $\Lambda_b$, which corresponds to momenta at which the
different lines cross each other. For N$^3$LO, $\Lambda_b$ was
estimated in Ref.~\cite{Epelbaum:2014efa} to be $\Lambda_b = 600\;$MeV for
$R=0.8-1.0\;$fm, $\Lambda_b = 500\;$MeV for $R=1.1\;$fm
and $\Lambda_b = 400\;$MeV for $R=1.2\;$fm. 

It should be understood that the proposed approach to uncertainty
quantification assumes the validity of Eq.~(\ref{Expan}), which relies on naive
dimensional analyis and does not explicitly account for near-threshold 
enhancement of the amplitude in the case of an unnaturally large
scattering length. We, however, expect this to be largely accounted
for implicitly through employing  in the error analysis the information about the actual
size of $\Delta X^{(i)}$. 

In the following, I will
apply the approach for error analysis outlined above to selected observables in
the NN system. I  emphasize that our method does, of course, not require
the knowledge of experimental data and is applicable to any observable of
interest and for any particular choice of the regulator $R$ since it
does not rely on cutoff variation. 

\begin{figure}[t]
\centering
\includegraphics[width=\textwidth,clip]{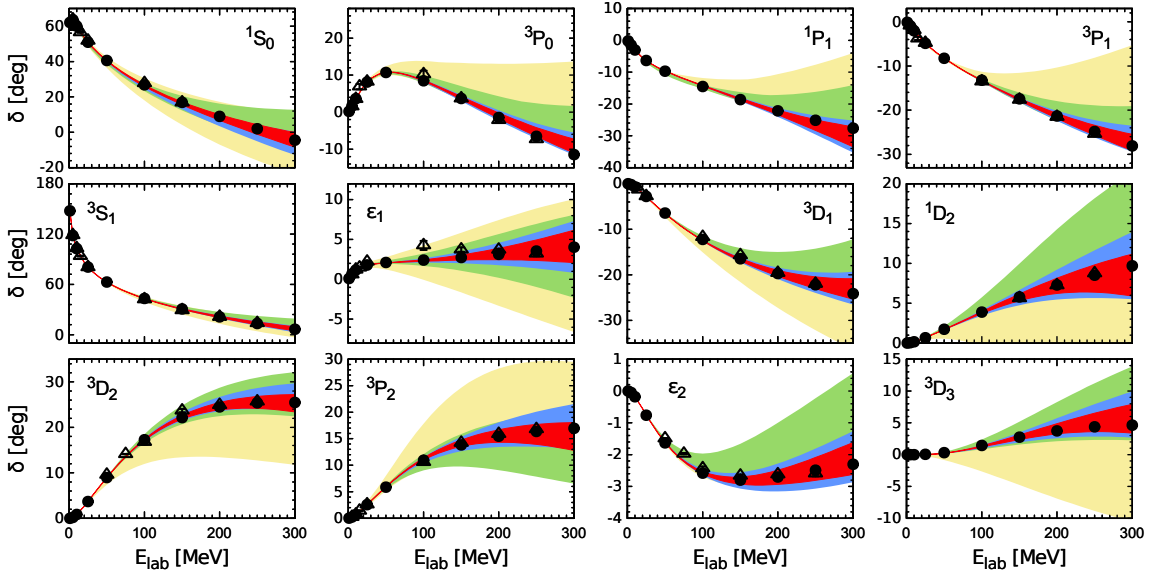}
\caption{np S-, P- and D- waves and the mixing angles
  $\epsilon_1$, $\epsilon_2$ at various chiral orders based on the cutoff of $R = 0.9\,$fm in comparison
  with the NPWA   \cite{Stoks:1993tb}  (solid dots) and the GWU
  single-energy PWA \cite{Arndt:1994br}
  (open triangles). The shaded bands 
show theoretical
  uncertainty at N$^4$LO (red), N$^3$LO (blue), 
  N$^2$LO (green) and NLO (yellow) estimated via 
  Eqs.~(\protect\ref{Err}) and (\protect\ref{ErrExplicit}).
}
\label{fig6}       
\end{figure}

\begin{figure}[t]
\centering
\includegraphics[width=\textwidth,clip]{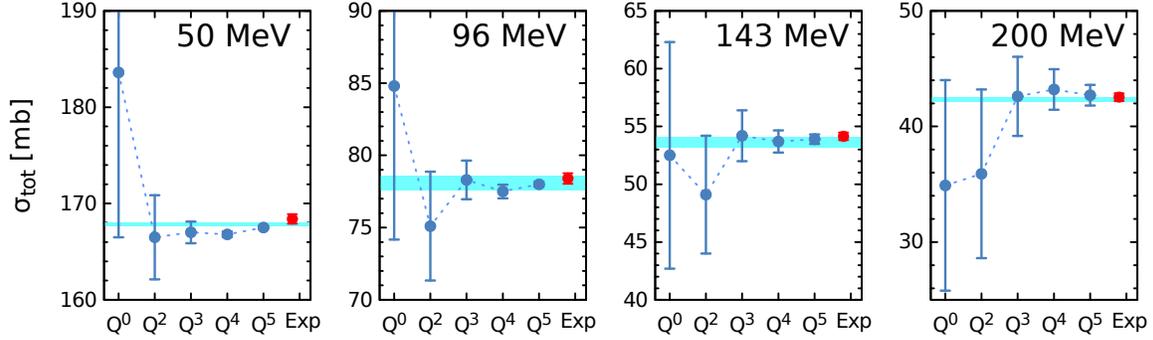}
\caption{Chiral expansion of the np total cross section at different energies based on $R=0.9\;$fm in comparison
  with experimental data of Ref.~\cite{Abfalterer:2001gw}.
The horizontal band shows the result of the NPWA.
}
\label{fig7}       
\end{figure}

\begin{figure}[h]
\centering
\includegraphics[width=0.8\textwidth,clip]{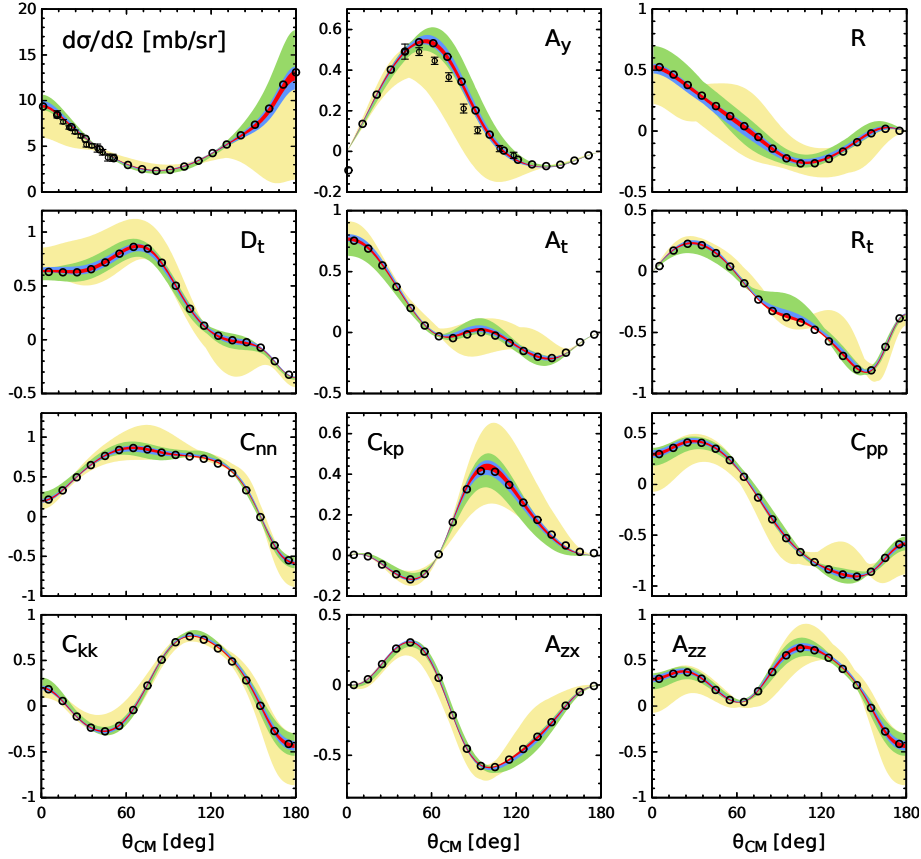}
\caption{Predictions for the np differential cross section $d \sigma/ d \Omega$,  the analyzing power $A_y$,
     the rotation parameter $R$, the polarization-transfer parameters
     $D_t$, $R_t$  and $A_t$ and the spin-correlation parameters
     $C_{nn}$, $C_{kp}$, $C_{pp}$, $C_{kk}$,  $A_{zx}$ and $A_{zz}$ 
at  $E_{\rm   lab} = 143\,$MeV calculated  up to N$^4$LO based on the
      cutoff of $R=0.9\,$fm.  
Data for the cross section are at $E_{\rm lab} = 142.8\,$MeV and taken from \cite{ber76} and for the analyzing power
from \cite{kuc61}.   For remaining notation see Fig.~\protect\ref{fig6}.
}
\label{fig9}       
\end{figure}

Fig.~\ref{fig6} shows our results for phase shifts calculated up to
N$^4$LO for $R=0.9\;$fm already depicted in Fig.~\ref{fig3}, which are now
furnished with the estimated theoretical uncertainties using
Eqs.~(\ref{Err}), (\ref{ErrExplicit}).  
The various bands result by adding/subtracting the estimated
theoretical uncertainty  to/from the calculated results. Similarly, we
show in Fig.~\ref{fig7} our predictions for the np total cross section 
at various energies using the same value of the regulator 
in comparison with the result of the NPWA. 
As in the case of phase shifts, one observes a very good convergence
of the chiral expansion and excellent agreement between the
theoretical predictions, NPWA and experimental data.  The convergence
appears to be very fast at the lowest considered energy and, as expected, slows
down at $E_{\rm lab}= 200\;$MeV, where the N$^4$LO predictions are, however,
still accurate within a few percent. 
Our quoted theoretical uncertainties for the total cross section and
the case of $R=0.9\;$fm were found in \cite{Furnstahl:2015rha} to
be consistent with the $68\%$ degree-of-belief intervals for EFT
predictions. 
As another application, we show in Fig.~\ref{fig9} our predictions for a selected set
of np scattering observables at $E_{\rm lab}= 143\;$MeV based on
$R=0.9\;$fm. 
In all cases, we observe excellent agreement with the
NPWA and confirm a good
convergence of the chiral expansion. More results for NN observables 
can be found in Refs.~\cite{Epelbaum:2014efa,Epelbaum:2014sza}. 

\begin{figure}[t]
\centering
\includegraphics[width=\textwidth,clip]{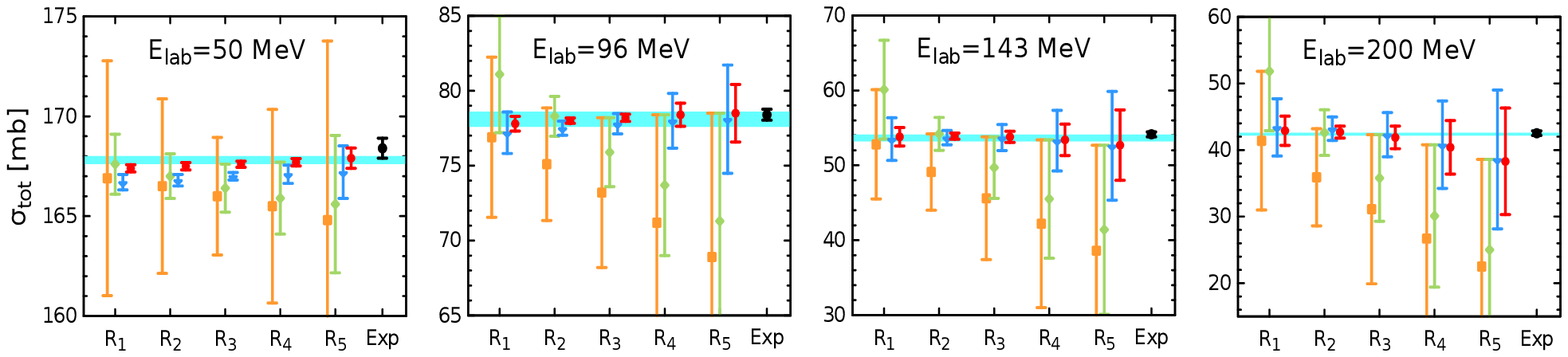}
\caption{Predictions for the np total cross section 
at NLO 
(filled orange squares), N$^2$LO (solid green diamonds), N$^3$LO
(filled blue triangles) and N$^4$LO (filled red circles) for different
choices of the cutoff, see Eq.~(\protect\ref{RegSp}) in comparison
with NPWA (the horizontal bands) and data of Ref.~\cite{Abfalterer:2001gw}.
}
\label{fig8}       
\end{figure}

As already advertised, the novel approach to uncertainty
quantification is not restricted to a particular choice of the
regulator. Carrying out the error analysis for calculations based on
different choices of $R$ thus provides a useful consistency check of the
method. In Fig.~\ref{fig8}, we show the results for the total cross
section at all orders starting from NLO and for all considered cutoff
choices. 
Within the quoted errors, the predictions based on different values of
$R$ agree with each other and the NPWA for all orders in the
chiral expansion. The accuracy of the predicted results
for the cross section shows the same dependence on the cutoff 
as the quality of the fits 
discussed in section \ref{PhS}. 

In Table \ref{tab_deut}, we list our results for the deuteron properties. 
\begin{table}[t]
\caption{Deuteron binding energy $B_d$ (in MeV), asymptotic $S$ state
  normalization $A_S$ (in fm$^{-1/2}$), asymptotic $D/S$ state ratio $\eta$, radius
  $r_d$ (in fm), quadrupole moment $Q$  (in fm$^2$) and the $D$-state
  probability $P_D$ (in $\%$)
based on the cutoff $R=0.9$ fm. 
Notice that $r_d$ and $Q$ are calculated
  without including 
exchange current contributions and
  relativistic corrections.
  References to experimental data/empirical values can be found in Ref.~\cite{Epelbaum:2014efa}.    
\label{tab_deut}}
\begin{tabularx}{\textwidth}{@{\extracolsep{\fill}}lllllll}
\hline
\noalign{\smallskip}
   &  LO & NLO  & N$^2$LO  & N$^3$LO  & N$^4$LO  & Empirical
\smallskip
 \\
\hline 
$B_d$ & 2.0235 & 2.1987 & 2.2311 & 2.2246$^\star$ & 2.2246$^\star$ & 2.224575(9) \\ 
$A_S$ & 0.8333  & 0.8772  & 0.8865  & 0.8845  & 0.8844  & 0.8846(9) \\ 
$\eta$ & 0.0212 & 0.0256 & 0.0256 & 0.0255 &  0.0255 &  0.0256(4)\\ 
$r_d$& 1.990 &  1.968 & 1.966 & 1.972 &  1.972 &  1.97535(85)\\ 
$Q$ & 0.230 & 0.273 & 0.270 & 0.271 &   0.271 &  0.2859(3)\\ 
$P_D$ & 2.54 & 4.73 & 4.50 & 4.19 &  4.29 &  \\
\hline
\multicolumn{7}{l}{$^\star$The deuteron binding energy has been
  taken as input in the fit.} 
\end{tabularx}
\end{table}
At the considered accuracy level, the chiral expansion is nearly
converged already at N$^3$LO except for $P_D$ which is not an
observable quantity.\footnote{$P_D = 5\% \pm 1\%$ has been used as an
  additional ``data'' point in the fits at N$^3$LO and N$^4$LO in order to stabilize
  the results, see Ref.~\cite{Epelbaum:2014efa} for more detail.}
The predicted values for $A_S$ and $\eta$ are in excellent agreement
with the empirical numbers. Using Eqs.~(\ref{Err}) and
(\ref{ErrExplicit}) and adopting $Q=M_\pi/\Lambda_b$, 
our predictions for $A_S$  at N$^4$LO is $A_S=0.8844 \pm 0.0002
\;$fm$^{-1/2}$ while the accuracy for $\eta$ is beyond the quoted
figures. For the $r_d$ and $Q$, our results are
incomplete as we do not include relativistic corrections and
meson-exchange current contributions. The estimated size of
these corrections is consistent with the deviation between our values
and the empirical numbers, see \cite{Epelbaum:2014efa} for an extended discussion.

\section{Beyond the two-nucleon system}
\label{sec4}

\begin{figure}[t]
\centering
\includegraphics[width=\textwidth,clip]{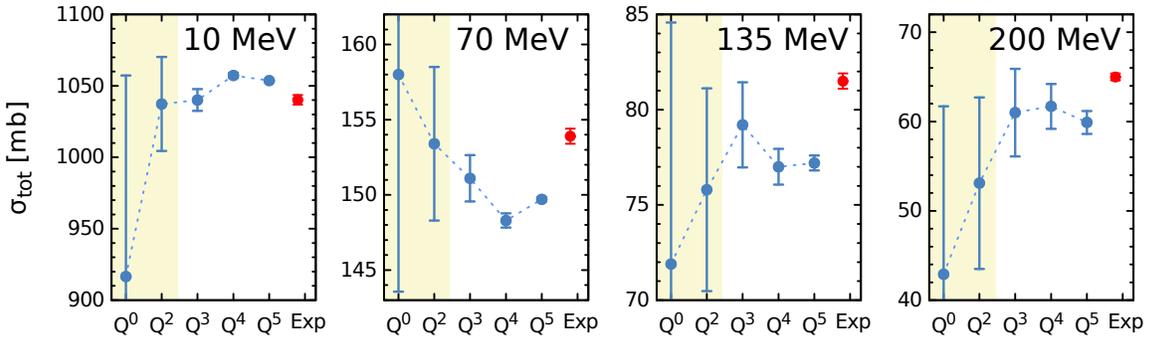}
\caption{Chiral expansion of the Nd total cross section at different
  energies based on the cutoff or $R=0.9\;$fm and using NN forces only
  in comparison
  with experimental data of Ref.~\cite{Abfalterer:2001gw}.
}
\label{fig10}       
\end{figure}

Having developed the new generation of NN potentials up to N$^4$LO and
the novel approach to uncertainty quantification, which has been
validated in the NN system, we are well prepared to test nuclear
chiral EFT in heavier systems and to systematically analyze the role
of the 3NF, which has been the subject of intense experimental 
research at FZ J\"ulich, GANIL, KVI, RIKEN, TUNL and other laboratories.  This is the main goal of the recently
formed Low Energy Nuclear Physics International Collaboration
(LENPIC). The numerical implementation of the 3NF regularized in the
same way as the NN potentials of
Refs.~\cite{Epelbaum:2014efa,Epelbaum:2014sza} is currently in
progress so that no results including the novel NN potentials and 3NF
are available at this stage. Still, as argued in \cite{Binder:2015mbz}, it is instructive to
analyze few-nucleon systems based on NN forces only. Consider, for
example, the nucleon-deuteron (Nd)  total cross section. Our
predictions for this observable at four different energies are
summarized  in Fig.~\ref{fig10} for the case of $R=0.9\;$fm. 
The light-shaded yellow area in the plots indicates that our results
are complete only up to NLO due to missing 3NF. 
One observes a significant discrepancy
between the theoretical predictions based on the NN forces only and
data, which provides clear evidence for missing 3NF contributions, 
cf.~Figs.~\ref{fig7} and \ref{fig10}. 
The estimated accuracy of the chiral EFT results at N$^4$LO suggests
that Nd scattering at intermediate energies will be a very
promising testing ground for the chiral 3NF. Furthermore, it is
comforting to see that the deviations between our N$^{3}$LO/N$^4$LO
predictions and experimental data are similar in size to the NLO
error bars, which give the estimated size of
N$^2$LO contributions. This is precisely the chiral order at which the
first nonvanishing 3NF appears according to Weinberg's power counting.  
Our results do not indicate the need to promote the short-range 3NF
operators to lower orders as suggested e.g.~in Ref.~\cite{Birse:2010fj}. 
For more examples of few-nucleon observables see
Ref.~\cite{Binder:2015mbz}.

\section{Summary and outlook}
\label{sec5}

In this talk I presented the new generation of  NN potentials derived
in chiral EFT up to N$^4$LO. We were able to significantly reduce
finite-cutoff artifacts by using an appropriate regularization in
coordinate space which maintains the analytic structure of the
amplitude.  The new potentials do not require the additional spectral
function regularization and employ the LECs
$c_i$,  $d_i$ and $e_i$ determined from $\pi N$ scattering without any
fine tuning. We found a clear evidence of the corresponding
parameter-free two-pion exchange contributions by observing an
improved description of NN phase shifts at N$^2$LO and N$^4$LO. 
Furthermore, a simple approach for estimating the theoretical
uncertainty in few-nucleon calculations from the truncation of the
chiral expansion, that does not rely on cutoff variation, was formulated
and validated in the NN system. 

Our work opens up new perspectives for precision ab initio
calculations in few- and many-nucleon systems and is especially
relevant for  ongoing efforts towards a quantitative understanding of
the structure of the 3NF in the framework of chiral EFT, see \cite{Binder:2015mbz}
for a first step along this line.


\begin{thebibliography}{99}
\bibitem{Weinberg:1990rz} 
  S.~Weinberg,
  Phys.\ Lett.\ B {\bf 251}, 288 (1990).

\bibitem{Weinberg:1991um} 
  S.~Weinberg,
  Nucl.\ Phys.\ B {\bf 363}, 3 (1991).

\bibitem{Epelbaum:2008ga} 
  E.~Epelbaum, H.~W.~Hammer and U.-G.~Mei{\ss}ner,
  Rev.\ Mod.\ Phys.\  {\bf 81}, 1773 (2009).

\bibitem{Machleidt:2011zz} 
  R.~Machleidt and D.~R.~Entem,
  Phys.\ Rept.\  {\bf 503}, 1 (2011).

\bibitem{KalantarNayestanaki:2011wz} 
  N.~Kalantar-Nayestanaki  {\it et al.},
  Rept.\ Prog.\ Phys.\  {\bf 75}, 016301 (2012).

\bibitem{Hammer:2012id} 
  H.~W.~Hammer, A.~Nogga and A.~Schwenk,
  Rev.\ Mod.\ Phys.\  {\bf 85}, 197 (2013).

\bibitem{Hagen:2012sh} 
  G.~Hagen  {\it et al.},
  Phys.\ Rev.\ Lett.\  {\bf 108}, 242501 (2012).

\bibitem{Jurgenson:2013yya} 
  E.~D.~Jurgenson {\it et al.},
  Phys.\ Rev.\ C {\bf 87}, no. 5, 054312 (2013).

\bibitem{Lovato:2013cua} 
  A.~Lovato {\it et al.},
  Phys.\ Rev.\ Lett.\  {\bf 111}, no. 9, 092501 (2013).

\bibitem{Hergert:2012nb} 
  H.~Hergert {\it et al.},
  Phys.\ Rev.\ C {\bf 87}, no. 3, 034307 (2013).

\bibitem{Soma:2012zd} 
  V.~Soma, C.~Barbieri, T.~Duguet,
  Phys.\ Rev.\ C {\bf 87}, no. 1, 011303 (2013).

\bibitem{Lee:2008fa} 
  D.~Lee,
  Prog.\ Part.\ Nucl.\ Phys.\  {\bf 63}, 117 (2009).

\bibitem{Epelbaum:2011md} 
  E.~Epelbaum {\it et al.},
  Phys.\ Rev.\ Lett.\  {\bf 106}, 192501 (2011).

\bibitem{Epelbaum:2012qn} 
  E.~Epelbaum  {\it et al.},
  Phys.\ Rev.\ Lett.\  {\bf 109}, 252501 (2012).

\bibitem{Epelbaum:2013paa} 
  E.~Epelbaum {\it et al.},
  Phys.\ Rev.\ Lett.\  {\bf 112}, no. 10, 102501 (2014).

\bibitem{Elhatisari:2015iga} 
  S.~Elhatisari  {\it et al.},
  arXiv:1506.03513 [nucl-th].

\bibitem{UGMproc} Ulf-G.~Mei{\ss}ner, {\it these proceedings}. 

\bibitem{Epelbaum:2014efa} 
  E.~Epelbaum, H.~Krebs and U.-G.~Mei{\ss}ner,
  Eur.\ Phys.\ J.\ A {\bf 51}, no. 5, 53 (2015).

\bibitem{Epelbaum:2014sza} 
  E.~Epelbaum, H.~Krebs and U.-G.~Mei{\ss}ner,
  Phys.\ Rev.\ Lett.\  {\bf 115}, no. 12, 122301 (2015).

\bibitem{Fettes:2000gb} 
  N.~Fettes  {\it et al.},
  Annals Phys.\  {\bf 283}, 273 (2000)
  [Annals Phys.\  {\bf 288}, 249 (2001)].


\bibitem{BeaneProc} S.~Beane, {\it these proceedings}. 

\bibitem{SchindlerProc} M.~Schindler, {\it these proceedings}. 

\bibitem{Hemmert:1997ye} 
  T.~R.~Hemmert, B.~R.~Holstein and J.~Kambor,
  J.\ Phys.\ G {\bf 24}, 1831 (1998).

\bibitem{Epelbaum:2007us} 
  E.~Epelbaum,
  Eur.\ Phys.\ J.\ A {\bf 34}, 197 (2007).

\bibitem{Weinberg:1992yk} 
  S.~Weinberg,
  Phys.\ Lett.\ B {\bf 295}, 114 (1992).

\bibitem{vanKolck:1994yi} 
  U.~van Kolck,
  Phys.\ Rev.\ C {\bf 49}, 2932 (1994).

\bibitem{Ordonez:1995rz} 
  C.~Ordonez, L.~Ray and U.~van Kolck,
  Phys.\ Rev.\ C {\bf 53}, 2086 (1996).

\bibitem{Pastore:2009is} 
  S.~Pastore  {\it et al.},
  Phys.\ Rev.\ C {\bf 80}, 034004 (2009).

\bibitem{Pastore:2011ip} 
  S.~Pastore  {\it et al.},
  Phys.\ Rev.\ C {\bf 84}, 024001 (2011).

\bibitem{Piarulli:2012bn} 
  M.~Piarulli  {\it et al.},
  Phys.\ Rev.\ C {\bf 87}, no. 1, 014006 (2013).

\bibitem{Baroni:2015uza} 
  A.~Baroni {\it et al.},
  arXiv:1509.07039 [nucl-th].

\bibitem{SchiProc}
R. Schiavilla, {\it these proceedings}.  

\bibitem{TMO}
M.~Taketani, S.~Machida and S.~Onuma, Prog. Theore. Phys. {\bf 7}
(1952) 45. 

\bibitem{Okubo:1954zz} 
  S.~Okubo,
  Prog.\ Theor.\ Phys.\  {\bf 12}, 603 (1954).

\bibitem{Epelbaum:1998ka} 
  E.~Epelbaum, W.~Gl\"ockle and U.-G.~Mei{\ss}ner,
  Nucl.\ Phys.\ A {\bf 637}, 107 (1998).

\bibitem{Epelbaum:2002gb} 
  E.~Epelbaum, U.-G.~Mei{\ss}ner and W.~Gl\"ockle,
  Nucl.\ Phys.\ A {\bf 714}, 535 (2003).

\bibitem{Epelbaum:2005pn} 
  E.~Epelbaum,
  Prog.\ Part.\ Nucl.\ Phys.\  {\bf 57}, 654 (2006).


\bibitem{Krebs:2004st} 
  H.~Krebs, V.~Bernard and U.-G.~Mei{\ss}ner,
  Annals Phys.\  {\bf 316}, 160 (2005).

\bibitem{Kolling:2009iq} 
  S.~K\"olling  {\it et al.},
  Phys.\ Rev.\ C {\bf 80}, 045502 (2009).

\bibitem{Kolling:2011mt} 
  S.~K\"olling {\it et al.},
  Phys.\ Rev.\ C {\bf 84}, 054008 (2011).

\bibitem{Kaiser:1997mw} 
  N.~Kaiser, R.~Brockmann and W.~Weise,
  Nucl.\ Phys.\ A {\bf 625}, 758 (1997).

\bibitem{Epelbaum:2006eu} 
  E.~Epelbaum,
  Phys.\ Lett.\ B {\bf 639}, 456 (2006).

\bibitem{Kaiser:1999ff} 
  N.~Kaiser,
  Phys.\ Rev.\ C {\bf 61}, 014003 (2000).

\bibitem{Kaiser:1999jg} 
  N.~Kaiser,
  Phys.\ Rev.\ C {\bf 62}, 024001 (2000).

\bibitem{Kaiser:2001pc} 
  N.~Kaiser,
  Phys.\ Rev.\ C {\bf 64}, 057001 (2001).

\bibitem{Kaiser:2001at} 
  N.~Kaiser,
  Phys.\ Rev.\ C {\bf 65}, 017001 (2002).

\bibitem{Entem:2003ft} 
  D.~R.~Entem and R.~Machleidt,
  Phys.\ Rev.\ C {\bf 68}, 041001 (2003).

\bibitem{Epelbaum:2004fk} 
  E.~Epelbaum, W.~Gl\"ockle and U.-G.~Mei{\ss}ner,
  Nucl.\ Phys.\ A {\bf 747}, 362 (2005).

\bibitem{Entem:2014msa} 
  D.~R.~Entem {\it et al.},
  Phys.\ Rev.\ C {\bf 91}, no. 1, 014002 (2015).

\bibitem{Entem:2015xwa} 
  D.~R.~Entem {\it et al.},
  arXiv:1505.03562 [nucl-th].

\bibitem{Epelbaum:2002vt} 
  E.~Epelbaum {\it et al.},
  Phys.\ Rev.\ C {\bf 66}, 064001 (2002).

\bibitem{Ishikawa:2007zz} 
  S.~Ishikawa and M.~R.~Robilotta,
  Phys.\ Rev.\ C {\bf 76}, 014006 (2007).

\bibitem{Bernard:2007sp} 
  V.~Bernard {\it et al.},
  Phys.\ Rev.\ C {\bf 77}, 064004 (2008).

\bibitem{Bernard:2011zr} 
  V.~Bernard {\it et al.},
  Phys.\ Rev.\ C {\bf 84}, 054001 (2011).

\bibitem{Krebs:2012yv} 
  H.~Krebs, A.~Gasparyan and E.~Epelbaum,
  Phys.\ Rev.\ C {\bf 85}, 054006 (2012).

\bibitem{Krebs:2013kha} 
  H.~Krebs, A.~Gasparyan and E.~Epelbaum,
  Phys.\ Rev.\ C {\bf 87}, no. 5, 054007 (2013).

\bibitem{Girlanda:2011fh} 
  L.~Girlanda, A.~Kievsky and M.~Viviani,
  Phys.\ Rev.\ C {\bf 84}, 014001 (2011).

\bibitem{KrebsProc} H.~Krebs, {\it these proceedings}. 

\bibitem{Kaplan:1998we} 
  D.~B.~Kaplan, M.~J.~Savage and M.~B.~Wise,
  Nucl.\ Phys.\ B {\bf 534}, 329 (1998).

\bibitem{Fleming:1999ee} 
  S.~Fleming, T.~Mehen and I.~W.~Stewart,
  Nucl.\ Phys.\ A {\bf 677}, 313 (2000).

\bibitem{Epelbaum:2012ua} 
  E.~Epelbaum and J.~Gegelia,
  Phys.\ Lett.\ B {\bf 716}, 338 (2012).

\bibitem{Kadyshevsky:1967rs} 
  V.~G.~Kadyshevsky,
  Nucl.\ Phys.\ B {\bf 6}, 125 (1968).

\bibitem{Epelbaum:2013ij} 
  E.~Epelbaum and J.~Gegelia,
  PoS CD {\bf 12}, 090 (2013).

\bibitem{Epelbaum:2013naa} 
  E.~Epelbaum {\it et al.},
  Eur.\ Phys.\ J.\ A {\bf 50}, 51 (2014).

\bibitem{Epelbaum:2015sha} 
  E.~Epelbaum {\it et al.},
  Eur.\ Phys.\ J.\ A {\bf 51}, no. 6, 71 (2015).

\bibitem{Nogga:2005hy} 
  A.~Nogga, R.~G.~E.~Timmermans and U.~van Kolck,
  Phys.\ Rev.\ C {\bf 72}, 054006 (2005).

\bibitem{PavonValderrama:2005wv} 
  M.~Pavon Valderrama and E.~Ruiz Arriola,
  Phys.\ Rev.\ C {\bf 74}, 054001 (2006).

\bibitem{Beane:2001bc} 
  S.~R.~Beane {\it et al.},
  Nucl.\ Phys.\ A {\bf 700}, 377 (2002).

\bibitem{Epelbaum:2009sd} 
  E.~Epelbaum and J.~Gegelia,
  Eur.\ Phys.\ J.\ A {\bf 41}, 341 (2009).

\bibitem{Lepage:1997cs} 
  G.~P.~Lepage,
  nucl-th/9706029.

\bibitem{Epelbaum:2006pt} 
  E.~Epelbaum and U.-G.~Mei{\ss}ner,
  Few Body Syst.\  {\bf 54}, 2175 (2013).

\bibitem{Baru:2012iv} 
  V.~Baru {\it et al.},
  Eur.\ Phys.\ J.\ A {\bf 48}, 69 (2012).

\bibitem{Gezerlis:2013ipa} 
  A.~Gezerlis {\it et al.},
  Phys.\ Rev.\ Lett.\  {\bf 111}, no. 3, 032501 (2013).

\bibitem{Piarulli:2014bda} 
  M.~Piarulli {\it et al.},
  Phys.\ Rev.\ C {\bf 91}, no. 2, 024003 (2015).


\bibitem{Epelbaum:2003gr} 
  E.~Epelbaum, W.~Gl\"ockle and U.-G.~Mei{\ss}ner,
  Eur.\ Phys.\ J.\ A {\bf 19}, 125 (2004).

\bibitem{Marji:2013uia} 
  E.~Marji {\it et al.},
  Phys.\ Rev.\ C {\bf 88}, no. 5, 054002 (2013).

\bibitem{Fettes:1998ud} 
  N.~Fettes, U.-G.~Mei{\ss}ner and S.~Steininger,
  Nucl.\ Phys.\ A {\bf 640}, 199 (1998).

\bibitem{Buettiker:1999ap} 
  P.~B\"uttiker and U.-G.~Mei{\ss}ner,
  Nucl.\ Phys.\ A {\bf 668}, 97 (2000).

\bibitem{Wendt:2014lja} 
  K.~A.~Wendt, B.~D.~Carlsson and A.~Ekstr\"om,
  arXiv:1410.0646 [nucl-th].

\bibitem{Hoferichter:2015tha} 
  M.~Hoferichter {\it et al.}, 
  arXiv:1507.07552 [nucl-th], to appear in Phys. Rev. Lett..

\bibitem{Hoferichter:2012wf} 
  M.~Hoferichter {\it et al.}, 
  JHEP {\bf 1206}, 063 (2012).

\bibitem{Long:2011xw} 
  B.~Long and C.~J.~Yang,
  Phys.\ Rev.\ C {\bf 85}, 034002 (2012).

\bibitem{Birse:2010fj} 
  M.~C.~Birse,
  Phil.\ Trans.\ Roy.\ Soc.\ Lond.\ A {\bf 369}, 2662 (2011).

\bibitem{Stoks:1993tb} 
  V.~G.~J.~Stoks {\it et al.},
  Phys.\ Rev.\ C {\bf 48}, 792 (1993).

\bibitem{Arndt:1994br} 
  R.~A.~Arndt, I.~I.~Strakovsky and R.~L.~Workman,
  Phys.\ Rev.\ C {\bf 50}, 2731 (1994).

\bibitem{Stoks:1994wp} 
  V.~G.~J.~Stoks  {\it et al.},
  Phys.\ Rev.\ C {\bf 49}, 2950 (1994).

\bibitem{Furnstahl:2014xsa} 
  R.~J.~Furnstahl, D.~R.~Phillips and S.~Wesolowski,
  J.\ Phys.\ G {\bf 42}, no. 3, 034028 (2015).


\bibitem{Ekstrom:2014dxa} 
  A.~Ekstr\"om {\it et al.},
  J.\ Phys.\ G {\bf 42}, no. 3, 034003 (2015).


\bibitem{GrieProc} H.~Grie{\ss}hammer, {\it these proceedings}. 

\bibitem{Abfalterer:2001gw} 
  W.~P.~Abfalterer  {\it et al.},
  Phys.\ Rev.\ C {\bf 63}, 044608 (2001).

\bibitem{Furnstahl:2015rha} 
  R.~J.~Furnstahl {\it et al.}, 
  Phys.\ Rev.\ C {\bf 92}, no. 2, 024005 (2015).

\bibitem{ber76} A.J. Bersbach et al., Phys. Rev. D {\bf 13} (1976) 535.
\bibitem{kuc61} A.F. Kuckes et al., Phys. Rev. {\bf 121} (1961) 1226.

\bibitem{Binder:2015mbz} 
  S.~Binder {\it et al.},
  arXiv:1505.07218 [nucl-th].

\end{thebibliography}
\end{document}